\documentclass[
preprint,
nofootinbib,
 amsmath,amssymb,
 aps,
 prd,
]{revtex4-2}

\usepackage{graphicx}
\usepackage{dcolumn}
\usepackage{bm}
\usepackage{hyperref}

\usepackage{color}

\begin{document}

\preprint{CERN-TH-2025-217}
\preprint{KCL-PH-TH/2025-17}
\preprint{Imperial-TP-2025-AR-1}

\title{Scalar effective potentials in de Sitter spacetime}
\author{Lucas Vicente García-Consuegra}
\email{lucas.vicente\_garcia-consuegra@kcl.ac.uk}
\affiliation{Department of Physics,\\
 King's College London,\\
Strand, London, WC2R 2LS, United Kingdom}

\author{Arttu Rajantie}%
\email{a.rajantie@imperial.ac.uk}
\affiliation{Theoretical Physics Department,
CERN,\\
1211 Geneva 23, Switzerland\\
}
\affiliation{Abdus Salam Centre for Theoretical Physics,\\
Imperial College London,\\
London, SW7 2AZ, United Kingdom}%

\date{\today}
\newpage

\begin{abstract}
We investigate two different definitions of a scalar field effective potential in quantum field theory in de Sitter spacetime: the standard textbook definition, and the constraint effective potential proposed by O'Raifeartaigh et al.~in 1986. While these definitions are equivalent in Minkowski spacetime, they differ significantly in de Sitter. We demonstrate this by computing them both explicitly at one-loop order in perturbation theory. It is well known that the perturbative expansion of the standard effective potential fails to converge for light fields. In contrast, the constraint effective potential does not suffer from this infrared problem, and it can therefore be computed using perturbation theory. We discuss the physical interpretation of the two effective potentials. In particular, we provide evidence supporting an earlier conjecture that the constraint effective potential is the correct one to use in the stochastic Starobinsky-Yokoyama theory.
\end{abstract}

\maketitle

\tableofcontents

\section{Introduction}

The effective potential is a powerful and widely used tool in quantum field theory. 
Incorporating quantum corrections to the classical potential of the scalar field, its minimum determines the vacuum state of the theory, and the power series expansion around the minimum gives the physical masses and couplings of the scalar fields in that vacuum state~\cite{Coleman:1973jx}. 
It was first introduced by Goldstone, Salam and Weinberg~\cite{Goldstone:1962es}, who defined it perturbatively as the sum of all one-particle irreducible vacuum Feynman diagrams and demonstrated that the vacuum state of the theory corresponds to its minimum. Shortly afterwards, a non-perturbative definition in terms of a Legendre transform of the generating functional was given by DeWitt~\cite{DeWitt:1964mxt} and Jona-Lasinio~\cite{Jona-Lasinio:1964zvf}, and developed further by Jackiw~\cite{Jackiw:1974cv}. This is the standard textbook definition of the effective potential, and we will refer to it as the \emph{standard effective potential}. 

For standard particle physics applications, it is sufficient to calculate the effective potential in Minkowski spacetime, but in cosmology the effects of spacetime curvature are important and have to be included. This has been done explicitly in de Sitter spacetime~\cite{Shore:1979as,Allen:1983dg}, which is a good approximation for the spacetime metric during inflation and which has a high level of symmetry, making the one-loop calculation tractable.

In cosmology, the effective potential is often used to describe the dynamics of the field in the (semi)classical approximation, for example when studying cosmological phase transitions~\cite{Athron:2023xlk}, vacuum stability~\cite{Markkanen:2018pdo}, or the evolution of the inflaton and other scalar fields during inflation~\cite{Guth:1982ec,Hawking:1982my,Bardeen:1983qw}. In practice, this is usually done by simply using the effective potential in the place of the classical potential at the level of the equations of motion, 
or by using the Starobinsky-Yokoyama stochastic theory~\cite{Starobinsky:1986fx,Starobinsky:1994bd} which incorporates quantum fluctuations through a stochastic noise term.
Either way, one cannot simply use the tree-level potential, because in a quantum field theory it is ultraviolet divergent (or cutoff dependent), and intuitively the effective potential appears to be the obvious choice.

There is, however, a serious problem with this, as one can see by considering the case of de Sitter spacetime. If one computes the effective potential for a light scalar field, with a mass less than the curvature scale, one finds that it is dominated by the long-wavelength infrared modes. This has two consequences: First, the quantum correction is large and therefore it needs to be computed accurately, rather than just naively replacing the classical potential with its intuitively obvious quantum analogue. And second, the contribution from each order in the loop expansion is larger than the previous one, and therefore perturbation theory breaks down.\footnote{In this paper, we do not consider the case of a free, massless, minimally coupled scalar field. That theory has a more serious infrared problem, as a result of which it has no de Sitter invariant vacuum state~\cite{Allen:1985ux, Allen:1987tz,Folacci:1992xc}} Various approximation and resummation techniques have been developed to cope with this \emph{infrared problem} at the technical level~\cite{Riotto:2008mv,Rajaraman:2010xd,Serreau:2011fu,Prokopec:2011ms,Arai:2011dd,Beneke:2012kn,Youssef:2013by,Arai:2013jna,LopezNacir:2013alw,Garbrecht:2014dca,Guilleux:2015pma,Moss:2016uix,Markkanen:2018bfx,Moreau:2018ena,Cespedes:2023aal, DiPietro:2023inn,Huenupi:2024ksc,Nath:2025vhg}.
In this paper, instead, we ask whether one should be actually calculating a different quantity altogether.

Indeed, it was recently noted in ref.~\cite{Camargo-Molina:2022paw} that the latter appears to be the case for the stochastic approximation. The authors calculated vacuum decay rates to one-loop order in quantum field theory and in the stochastic approximation and found that they agree if, instead of the standard effective potential, the calculation in the stochastic theory is carried out using the \emph{constraint effective potential}, which was introduced in 1986 by O'Raifeartaigh, Wipf and Yoneyama~\cite{ORaifeartaigh:1986axd}.

The original motivation for the constraint effective potential was that it is more practical for a numerical evaluation using Monte Carlo simulations than the standard definition. Instead of a Legendre transform, it is defined by introducing a delta function directly in the path integral. In the infinite volume limit, both definitions give the same function, and therefore the choice between them is purely a matter of convenience. However, in a finite volume they are different, and in practice this applies to de Sitter spacetime, too.

This situation with these two effective potentials has a direct analogy with undergraduate statistical physics. Let us consider an open thermodynamic system, in which the energy $E$ and particle number $N$ can vary. It is convenient to write $N=Vn$, where $n$ is the particle density and $V$ is the volume. We can introduce the chemical potential $\mu$ and the grand/Landau potential
\begin{equation}
\label{equ:grandpotential}
    \Omega(T,\mu)
    =-k_BT \ln \sum_i e^{-\frac{(E_i-\mu V n_i)}{k_BT}},
\end{equation}
where the sum is over all microstates, labelled by $i$.
The average number density of particles is then given by
\begin{equation}
    \langle n\rangle=-\frac{1}{V}\frac{\partial\Omega}{\partial \mu}.
\end{equation}
One can then take the Legendre transform with respect to the chemical potential $\mu$ to obtain the Helmholtz free energy
\begin{equation}
\label{equ:F1}
F(T,n)=\Omega(T,\mu)+\mu Vn.
\end{equation}
This is analogous to the definition of the standard effective potential. The free energy (\ref{equ:F1}) has the property that its minimum determines the ensemble average of $n$, in the same way as the minimum of the standard effective potential 
determines the vacuum expectation value of the scalar field in quantum field theory.

In statistical physics, one can also define the Helmholtz free energy directly using the canonical ensemble, i.e., by fixing the number density of particles $n$. This way, one writes it as
\begin{equation}
\label{equ:F2}
    \tilde{F}(T,n)=-k_BT\ln \sum_j e^{-\frac{E_j}{k_BT}},
\end{equation}
where the sum is over all microstates with number density $n$, labelled by $j$, and we use the tilde ($\sim$) to distinguish it from $F(T,n)$ defined in eq.~(\ref{equ:F1}). This is analogous to the definition of the constrained effective potential.

In the thermodynamic limit,
the two free energies (\ref{equ:F1}) and (\ref{equ:F2}) are equal, but in a finite volume they are not. The reason for the difference is not that eqs.~(\ref{equ:F1}) and (\ref{equ:F2}) describe different physics, or that one is more correct than the other. Instead, they are simply two different functions of state describing the same physical system. For example, the minimum of $F(T,n)$ gives the expectation value of $n$, and the minimum of $\tilde{F}(T,n)$ gives the most likely value of $n$, and in a finite volume, these two quantities are not equal in general. 
More generally, $\tilde{F}(T,N)$ gives the probability distribution $p(T,n)$ of the number density,
\begin{equation}
\label{equ:pF}
    p(T,n)\propto \exp\left(-\frac{\tilde{F}(T,n)}{k_BT}\right).
\end{equation}

The aim of this paper is to compute the standard and constraint effective potentials explicitly to one-loop order for a real scalar field in de Sitter and discuss their differences. In particular, we will find that the constraint effective potential does not suffer from the same infrared problem as the standard effective potential. We will also provide evidence to support the conjecture in ref.~\cite{Camargo-Molina:2022paw} that the constraint effective potential is what one should use in the Starobinsky-Yokoyama stochastic theory~\cite{Starobinsky:1986fx,Starobinsky:1994bd}.

The paper is organised as follows. Section \ref{Sec:QFTdS} presents the Lorentzian and Euclidean actions for an interacting scalar field on de Sitter space and sets out the renormalisation scheme and counterterms. Sections \ref{Sec:sEP} and \ref{Sec:cEP} contain the one-loop computations of the standard and constraint effective potentials for fields in the trivial and fundamental representations of the orthogonal group, together with the required regularisation steps. We extract the effective parameters through power-series expansions in three physically relevant regimes: heavy, near conformal, and light. Section \ref{Sec:Discussion} discusses the implications of these results, with emphasis on the infrared behaviour of light fields and its relation to the stochastic description. Section \ref{Sec:Conclusions} summarises our findings and outlines possible extensions. Appendix \ref{Sec:Appendix_A} records the derivation of the integration constant used in the regularisation procedure.

\section{Quantum Field Theory in de Sitter spacetime\label{Sec:QFTdS}}

We consider a scalar field theory in a fixed de Sitter spacetime, whose metric in global coordinates is
\begin{equation}
\label{equ:deSitter}
    ds^{2} =g_{\mu\nu} dx^\mu dx^\nu= -dt^{2} + \frac{\cosh^{2}(Ht)}{H^{2}}\, d\Omega_{3}^{2},
\end{equation}
where $d\Omega_{3}^{2}$ denotes the line element on the unit three-sphere and $H$ is a real, constant parameter. This spacetime is maximally symmetric with curvature $R = 12H^{2}$ and arises as the solution to Einstein’s equations sourced by a constant positive energy density. Because of its high degree of symmetry, it is convenient for calculations, but it is also a good approximation of the inflationary spacetime in cosmology when $H$ is identified with the Hubble rate. In this work we treat the geometry as fixed, so the dynamics involve only the matter sector, and gravitational degrees of freedom are not included. 

In terms of the bare field $\phi_B$, the action is
\begin{equation}
    S[\phi_B]=\int d^4x\sqrt{-g}\, \left(-\frac{1}{2}g^{\mu\nu}\partial_\mu\phi_B\partial_\nu\phi_B-V_B(\phi_B)\right),
\label{Eq:Scalar_Action}
\end{equation}
where $g=\det g_{\mu\nu}$ and $V_B(\phi_B)$ is the bare potential. For concreteness, in this work we study
\begin{equation}
    V_B(\phi_B)=\frac{1}{2}(m_B^2+\xi_B R)\phi_B^2+\frac{1}{4}\lambda_B\phi_B^4,
\label{Eq:Classical_Potential}
\end{equation}
where $m_B$, $\xi_B$, and $\lambda_B$ are the bare mass parameter, non-minimal gravitational coupling, and self-coupling constant, respectively. Throughout, we will restrict our analysis to the weakly coupled regime $\lambda \ll 1$ and treat the theory perturbatively around free-field solutions.

The bare quantities, denoted by a subscript $B$, are related to their renormalised counterparts (no subscript) and counter terms through the standard field and parameter redefinitions
\begin{align}
    \label{Eq:RE}
        \phi_{B}&=\sqrt{Z_\phi}\,\phi,
        \nonumber\\
        m_B^2&=Z_\phi^{-1} (m^2+\delta m^2),
        \nonumber\\
         \xi_B&= Z_\phi^{-1}(\xi+\delta \xi),
        \nonumber\\
          \lambda_B&= Z_\phi^{-2}\left(\lambda+\delta\lambda\right).
\end{align}
We use dimensional regularisation and adopt the modified minimal subtraction ($\overline{\text{MS}}$) scheme. The counterterms are independent of the spacetime metric and therefore the same as in Minkowski space. For this theory, they are known in the literature (see refs.~\cite{Birrell:1982ix, Parker:2009uva}) and read
\begin{align}
        Z_\phi&=1+\mathcal{O}(\lambda^2), \nonumber
        \\
        \delta{m^2}&=\frac{3\lambda m^2}{16\pi^2}\left[\frac{2}{\varepsilon}-\gamma+\ln(4\pi)\right]+\mathcal{O}(\lambda^2)\nonumber,
        \\
        \delta{\xi}&=\left(\xi-\frac{1}{6}\right)\frac{3\lambda}{16\pi^2}\left[\frac{2}{\varepsilon}-\gamma+\ln(4\pi)\right]+\mathcal{O}(\lambda^2)\nonumber,
        \\
        \delta{\lambda}&=\frac{9\lambda^2}{16\pi^2}\left[\frac{2}{\varepsilon}-\gamma+\ln(4\pi)\right]+\mathcal{O}(\lambda^3).
\label{Eq:Counterterms}
\end{align}

Because the Ricci scalar $R$ is constant in de Sitter, we can combine the two quadratic terms in (\ref{Eq:Classical_Potential}) into one parameter,
\begin{equation}
\label{Eq:capitalM2}
    M^2=m^2+12\xi H^2.
\end{equation}
This quantity uniquely characterises the two-point behaviour of the theory, with the conformal point given by $M^2=2H^2$, irrespectively of the individual values of $m^2$ and $\xi$. Its corresponding counter-term is given by
\begin{equation}
    \label{Eq:capitaldeltaM2}
            \delta{M^2}=\frac{3\lambda}{16\pi^2}\left(M^2-2H^2\right)\left[\frac{2}{\varepsilon}-\gamma+\ln(4\pi)\right].
\end{equation}

For later convenience, we express the bare potential (\ref{Eq:Classical_Potential}) as a sum of the tree-level renormalised potential $V(\phi)$ and the counterterms $\delta V(\phi)$ as
\begin{equation}
V_B(\phi_B)=V(\phi)+\delta V(\phi),
\end{equation}
where
\begin{equation}
    V(\phi)=\frac{1}{2}M^2\phi^2+\frac14\lambda\phi^4.
\end{equation}

Because the spectrum of the  d'Alembertian on de Sitter space is unbounded from below, path integration is not well defined in Lorentzian signature \cite{Allen:1983dg}. To obtain a fluctuation operator with a positive-definite, discrete spectrum, we perform the analytic continuation $t \mapsto i\tau$. This takes the metric~(\ref{equ:deSitter}) to
\begin{equation}
\label{equ:spheremetric}
ds^2=d\Omega_4^2=d\tau^2 +\frac{\cos^2H\tau}{H^2}d\Omega_3^2,
\end{equation}
where the time coordinate has become compactified with a period of $2\pi/H$. This is the metric of a four-sphere, $S^4$, of radius $1/H$, which has finite volume
\begin{equation}
\label{equ:Omega4}
    \Omega_4=\int d^4x\sqrt{g}=\frac{8\pi^2}{3H^4}.
\end{equation}

In this Euclidean space the kinetic operator becomes strongly elliptic, and thus the spectrum of the theory is controlled. 
In terms of the renormalised field, the action is
\begin{align}
    S_{\text{E}}[\phi]&=
    \int d^4x\sqrt{g}\, \left(\frac{1}{2}g^{\mu\nu}\partial_\mu\phi_B\partial_\nu\phi_B+\frac{1}{2}M_B^2\phi_B^2+\frac{1}{4}\lambda_B\phi_B^4\right)\nonumber\\
    &=\int d^4x\sqrt{g}\, \left[\frac{1}{2}Z_\phi g^{\mu\nu}\partial_\mu\phi\partial_\nu\phi
    +\frac{1}{2}\left(M^2+\delta M^2\right)\phi^2
    +\frac{1}{4}\left(\lambda+\delta\lambda\right)\phi^4
    \right],
    \label{equ:S4action}
\end{align}
where the subscript ${\text{E}}$ indicates Euclidean signature.

\section{Standard effective potential\label{Sec:sEP}}
\subsection{Definition\label{Sec:EffPots_sEP}}

In quantum field theory, the standard definition~\cite{Jackiw:1974cv} of the effective potential follows the same steps as the first definition (\ref{equ:F1}) of the Helmholtz free energy density in thermodynamics. We start by defining the partition function for the scalar field $\phi$ in the presence of an external source $J$. In textbooks, it is often taken to be a functional because the aim is to define the one-particle irreducible effective action. However, because we are only interested in the effective potential, we only need the partition function for a constant $J$,
\begin{equation}
\mathcal{Z}_\mathcal{S}(J)=\int \mathcal{D}\phi\, \exp\left[-S_{\text{E}}[\phi]+J\int d^4x\sqrt{g}\, \phi(x)\right].
\label{Eq:Standard_Partition_Function}
\end{equation}

In analogy with the grand potential (\ref{equ:grandpotential}), we define
\begin{equation}
\label{equ:WSJ}
\mathcal{W}(J) = -\frac{1}{\Omega_4}\ln \mathcal{Z}_{\mathcal{S}}(J).
\end{equation}
The expectation value of the field in the presence of the external source $J$ is
\begin{equation}
\label{Eq:Expectation_val}
\langle \phi\rangle_J=- 
\frac{d \mathcal{W}(J)}{d J},
\end{equation}
where $\Omega_4$ is the volume (\ref{equ:Omega4}).

Now, we can define the standard effective potential as the Legendre transform of $\mathcal{W}(J)$,
\begin{equation}
    \label{equ:VSdef}
    \mathcal{V}_\mathcal{S}(\bar\phi)=\mathcal{W}(J)+J\bar\phi,
\end{equation}
where $J$ is the solution of the equation
\begin{equation}
    -\frac{d \mathcal{W}(J)}{d J}=\bar\phi.
\end{equation}
It follows that the standard effective potential (\ref{equ:VSdef}) satisfies
\begin{equation}
    \frac{d\mathcal{V}_\mathcal{S}(\bar\phi)}{d\bar\phi}=0\quad\text{for}\quad \bar\phi=\langle\phi\rangle.
\end{equation}
In other words, the expectation value of the field (for $J=0$) is given by the minimum of $\mathcal{V}_\mathcal{S}(\bar\phi)$.

\subsection{Real scalar singlet}
\subsubsection{One-loop calculation}

Let us first compute the standard effective potential $\mathcal{V}_\mathcal{S}(\bar\phi)$ for a single real scalar field $\phi$, following refs.~\cite{Shore:1979as,Allen:1983dg}.
To evaluate the path integral (\ref{Eq:Standard_Partition_Function}), we expand the field about the classical solution in the presence of the external source $J$, which we denote by $\bar\phi$,
\begin{equation}
\phi(x)=\bar\phi+\varphi(x),
\label{Eq:ModeEXP_sEP}
\end{equation}
where $\varphi\sim\mathcal{O}(\hbar^{1/2})$. This affine change of variables carries no Jacobian, i.e., $\mathcal{D}\phi \mapsto \mathcal{D}\varphi$, and gives the quadratic action
\begin{align}
    \begin{split}
        S_{\text{E}}[\phi] &= S_{\text{E}}[\bar{\phi}] 
    + \frac{1}{2}\int d^4x\sqrt{g}\, \varphi\,\hat{O}(\bar\phi)\,\varphi+\mathcal{O}(\hbar^2),
    \end{split}
\label{Eq:ExpandedS_sEP}
\end{align}
where
we have defined the four-sphere Klein-Gordon operator,
\begin{equation}
    \label{Eq:EKG}
    \hat{O}(\bar\phi)=\Delta_{S^4}+V''(\bar\phi)=\Delta_{S^4}+M^2+3\lambda\bar{\phi}^2,
\end{equation}
with $\Delta_{\text{S}^4}=-\frac{1}{\sqrt{g}}\partial_\mu(\sqrt{g}g^{\mu\nu}\partial_\nu)$ as the four-sphere Laplacian;
and used the fact that the four-sphere is boundary-less to drop surface integrals after integration by parts.

From eq.~(\ref{equ:WSJ}), we obtain
\begin{align}
      \mathcal{W}(J)&= V(\bar\phi)+\delta V(\bar\phi)
     -\frac{1}{\Omega_4}\ln\left[\int \mathcal{D}\varphi\, \exp\left(-\frac{1}{2}\int d^4x\sqrt{g}\, \varphi\hat{O}(\bar\phi)\varphi\right)+\mathcal{O}(\hbar^2)\right] -  J\bar\phi
     \nonumber\\
     &= V(\bar\phi)+\delta V(\bar\phi)
     +\frac{1}{2\Omega_4}\ln\det\left(\frac{\hat{O}(\bar\phi)}{H^2}\right) +\text{constant} +\mathcal{O}(\hbar^2) -  J\bar\phi
     ,
\label{Eq:Connected_sEP}
\end{align}
where we have introduced the denominator $H^2$ inside the determinant to make it dimensionless, and the term denoted by ``constant" includes all field-independent contributions. The effective potential is then given by the Legendre transform (\ref{equ:VSdef}),
\begin{equation}
    \mathcal{V}_{\mathcal{S}}(\bar\phi)= V(\bar\phi)+\delta V(\bar\phi)+\frac{1}{2\Omega_4}\ln\det\left({\frac{\hat{O}(\bar\phi)}{H^2}}\right)+\text{constant}+\mathcal{O}(\hbar^2).    
\label{Eq:BareVsEP_sEP_singlet}
\end{equation}

Because $\hat{O}(\bar\phi)$ is an elliptic and self-adjoint operator on a compact manifold $S^4$, it has a complete set of eigenfunctions $\{\varphi_n\}$ and discrete, positive semi-definite eigenvalues $\{\omega_n^2\}$~\cite{GibbonsHawking_1},
\begin{equation}
\label{equ:eveq}
\hat{O}(\bar\phi)\varphi_n=\omega_n^2\varphi_n,\quad n\ge 0.    
\end{equation}
We can use 
them
to write
\begin{equation}
    \ln\det \left(\frac{\hat{O}(\bar\phi)}{H^2} \right)
    =\ln\left[\prod_{n=0}^\infty\left(\frac{\omega_n^2}{H^2}\right)^{d_n}\right]=\sum_{n=0}^\infty d_n\ln \frac{\omega^2_n}{H^2}, 
\label{Eq:LogDet_Linear_operator}
\end{equation}
where $d_n$ is the degeneracy of the $n$-th eigenvalue. This sum is divergent, and as discussed in section~\ref{Sec:QFTdS}, we regularise it using dimensional regularisation, by analytically continuing to $d=4-\varepsilon$ dimensions with $\varepsilon\ll1$.

In $d$ dimensions, the eigenfunctions $\varphi_n$ of $\hat{O}(\bar\phi)$ are the $d$-dimensional spherical harmonics. The eigenvalues $\omega_n^2$ and their corresponding degeneracies $d_n$ are~\cite{Shore:1979as,Allen:1983dg}
\begin{align}
\label{Eq:SpectrumLaplacian}
        \omega_{n}^2 &=H^{2}(n^2+(d-1)n)+V''(\bar\phi)=H^{2}\left(n+\frac{d-1}{2}+\nu_d\right)\left(n+\frac{d-1}{2}-\nu_d\right),
        \nonumber\\
        d_n&=\frac{(2n+d-1)\Gamma(n+d-1)}{\Gamma(n+1)\Gamma(d)},
\end{align}
where $\nu_d$ is the de Sitter mass parameter in $d$ dimensions,
\begin{equation}
    \nu_d{(\bar\phi)}=\sqrt{\frac{(d-1)^2}{4}-\frac{V''(\bar\phi)}{H^{2}}}.
\end{equation}
The lowest eigenfunction is a constant $\varphi_0(x)=1$, and its corresponding eigenvalue is simply $\omega_0^2=V''(\bar\phi)$, whereas all other eigenvalues include a term proportional to $H^2$.

To compute eq.~(\ref{Eq:LogDet_Linear_operator}) using dimensional regularisation, we define the function
\begin{align}
\label{Eq:I_1}
    \mathcal{F}_d(\nu_d)&=\frac{\mu^{4-d}}{2\Omega_d}\ln\det\left({\frac{\hat{O}(\bar\phi)}{H^2}}\right)\nonumber \\
    &=\frac{\mu^{4-d}}{2\Omega_d}\sum_{n=0}^\infty \frac{(2n+d-1)\Gamma(n+d-1)}{\Gamma(n+1)\Gamma(d)}\ln\left[\left(n+\frac{d-1}{2}+\nu_d\right)\left(n+\frac{d-1}{2}-\nu_d\right)\right],
\end{align}
where
\begin{equation}
    \Omega_d=\frac{2\pi^{\frac{d+1}{2}}}{ \Gamma\left(\frac{d+1}{2}\right)}\frac{1}{H^d}
\end{equation}
is the volume of a $d$-sphere, and we have introduced the renormalisation scale $\mu^{4-d}$ to keep the expression consistent across dimensions. To evaluate the series, we write
\begin{equation}
     \mathcal{F}_d(\nu_d)=\int d\nu_d\, \frac{d  \mathcal{F}_d}{d \nu_d}.
\label{Eq:FormalIVP}
\end{equation}

We begin by differentiating eq.~(\ref{Eq:I_1}) with respect to $\nu_d^2$,
\begin{align}
     \frac{d  \mathcal{F}_d}{d \nu_d^2} &=\frac{H^d \mu^{4-d}\ \Gamma\left(\frac{d+1}{2}\right)}{4\pi^{\frac{d+1}{2}}\Gamma(d)} \sum_{n=0}^\infty \frac{(2n+d-1)\Gamma(n+d-1)}{\Gamma(n+1)}\frac{1}{n^2+(d-1)n+\frac{(d-1)^2}{4}-\nu_d^2}\nonumber\\
     &=\frac{H^d \mu^{4-d}\ \Gamma\left(\frac{d+1}{2}\right)}{4\pi^{\frac{d+1}{2}}\Gamma(d)}\cos(\pi\nu_d)\csc\left(\frac{\pi d}{2}\right)\Gamma\left[\frac{(d-1)}{2}+\nu_d\right]\Gamma\left[\frac{(d-1)}{2}-\nu_d\right],
\end{align}
where to arrive at the last line we have performed the summation symbolically \cite{DiPietro:2023inn}.

Next, we deform away from $4$ spacetime dimensions by setting $d=4-\varepsilon$ and use the chain rule to re-write the derivative above in terms of $\nu_d$. This leads to 
\begin{align}
        \frac{d  \mathcal{F}_d}{d\nu_d }&=\frac{H^4 (4\nu_d^3-\nu_d)}{64\pi^2}\left[\frac{2}{\varepsilon}+\ln\left(\frac{4\pi\mu^2}{H^2}\right)-\gamma+1-\psi^{(0)}\left(\frac{3}{2}+\nu_d\right)-\psi^{(0)}\left(\frac{3}{2}-\nu_d\right)\right],
    \label{Eq:dIdn_1}
\end{align}
where $\gamma$ is the Euler-Mascheroni constant and $\psi^{(n)}(z)$ is the $n$-th polygamma function~\cite{Adamchik1998}. The integral (\ref{Eq:FormalIVP}) can now be evaluated analytically.

To take the $d\rightarrow 4-\varepsilon$ limit, we express $\nu_d$ in terms of the four-dimensional de Sitter mass parameter $\nu=\nu_4$ as
\begin{equation}
\label{equ:nuvsnud}
    \nu_{d}=\nu-\frac{3\varepsilon}{4\nu}+\mathcal{O}(\varepsilon^2),
\end{equation}
and write the four-dimensional dimensionally regularised one-loop term $\mathcal{F}(\nu)$ as
\begin{align}
         \mathcal{F}(\nu)&=\lim_{\varepsilon\rightarrow 0}
         \mathcal{F}_{4-\varepsilon}\left(\nu-\frac{3\varepsilon}{4\nu}\right)
         \nonumber\\
         &= -\frac{H^4}{64\pi^2}
        \left\{ \left(\nu^4-\frac{\nu^2}{2}\right)
        \left(\frac{2}{\varepsilon}
        +\ln\frac{4\pi^2\mu^2}{H^2}-\gamma+1\right)
        +6\nu^2\right.\nonumber\\
        &\hspace{2cm}-\left(4\nu^{3}-\nu\right)\left[\ln\!\Gamma\left(\frac{3}{2}+{\nu}\right)-\ln\!\Gamma\left(\frac{3}{2}-{\nu}\right)\right]\nonumber\\
        &\hspace{2cm}+\left(12\nu^2-1\right)\left[\psi^{(-2)}\left(\frac{3}{2}+{\nu}\right)+\psi^{(-2)}\left(\frac{3}{2}-{\nu}\right)\right]\nonumber\\
         &\hspace{2cm}-24\nu\left[\psi^{(-3)}\left(\frac{3}{2}+{\nu}\right)-\psi^{(-3)}\left(\frac{3}{2}-{\nu}\right)\right]\nonumber\\
         &\left.\hspace{2cm}+24\left[\psi^{(-4)}\left(\frac{3}{2}+{\nu}\right)+\psi^{(-4)}\left(\frac{3}{2}-{\nu}\right)\right]\right\}+\tilde{\mathcal{F}},
\label{Eq:F_final}
\end{align}
where the value of $\tilde{\mathcal{F}}$ is given in eq.~(\ref{equ:F0}).

Combining eqs.~(\ref{Eq:BareVsEP_sEP_singlet}) and (\ref{Eq:F_final}), we arrive at the full one-loop expression for the standard effective potential in de Sitter spacetime,\footnote{If we had omitted the step (\ref{equ:nuvsnud}), as is sometimes done in the literature, the expression would be identical except that the term $+6\nu^2(\bar\phi)$ on the second line would be absent. At least for the purposes of this calculation, this difference can be absorbed in the definition of the non-minimal coupling $\xi$.}
\begin{align}
\label{equ:fullpot}
\mathcal{V}_\mathcal{S}(\bar\phi)
    &=V(\bar\phi)+\delta V(\bar\phi)
    \nonumber\\
    &
    \quad-\frac{H^4}{64\pi^2}
        \left\{ \left(\nu^4{(\bar\phi)}-\frac{\nu^2{(\bar\phi)}}{2}\right)
        \left(\frac{2}{\varepsilon}
        +\ln\frac{4\pi\mu^2}{H^2}
        -\gamma+1\right)+6\nu^2{(\bar\phi)}
        \right.
        \nonumber\\
        &\quad\quad\quad\quad
        -\left(4\nu^2{(\bar\phi)}-\nu{(\bar\phi)}\right)\left[\ln\!\Gamma\left(\frac{3}{2}+{\nu{(\bar\phi)}}\right)-\ln\!\Gamma\left(\frac{3}{2}-{\nu{(\bar\phi)}}\right)\right]
        \nonumber\\
        &\quad\quad\quad\quad
        +\left(12\nu^2{(\bar\phi)}-1\right)\left[\psi^{(-2)}\left(\frac{3}{2}+{\nu{(\bar\phi)}}\right)+\psi^{(-2)}\left(\frac{3}{2}-{\nu{(\bar\phi)}}\right)\right]
        \nonumber\\
         &\quad\quad\quad\quad
         -24\nu{(\bar\phi)}\left[\psi^{(-3)}\left(\frac{3}{2}+{\nu{(\bar\phi)}}\right)-\psi^{(-3)}\left(\frac{3}{2}-{\nu{(\bar\phi)}}\right)\right]
         \nonumber\\
         &\left.\quad\quad\quad\quad
         +24\left[\psi^{(-4)}\left(\frac{3}{2}+{\nu{(\bar\phi)}}\right)+\psi^{(-4)}\left(\frac{3}{2}-{\nu{(\bar\phi)}}\right)\right]\right\}
         +\text{constant}+\mathcal{O}(\hbar^2),
\end{align}
where
\begin{equation}
\label{equ:nuofphi}
    \nu(\bar\phi)=\sqrt{\frac{9}{4}-\frac{V''(\bar\phi)}{H^2}}=\sqrt{\frac{9}{4}-\frac{M^2+3\lambda\bar\phi^2}{H^2}}.
\end{equation}
The $1/\varepsilon$ pole cancels against the counterterms $\delta V(\bar\phi)$, so that the full expression is finite in the $\varepsilon\rightarrow 0$ limit.

\subsubsection{Power series expansions}
\label{sec:powerseries}
While eq.~(\ref{equ:fullpot}) is the complete one-loop result for a single real scalar field, it is somewhat opaque because of the implicit field dependence through $\nu(\bar\phi)$. Therefore we expand it as a power series in the field $\bar\phi$
and calculate the coefficients up to sextic order in the background field,
\begin{equation}
    \label{equ:VSexp}
    \mathcal{V}_\mathcal{S}(\bar\phi)=\frac{1}{2} M^2_\mathcal{S}\bar\phi^2+\frac{1}{4}\lambda_\mathcal{S}\bar\phi^4+\frac{1}{6}\kappa_\mathcal{S}\bar\phi^6+\mathcal{O}(\bar\phi^8),
\end{equation}
for the theory (\ref{Eq:Classical_Potential}). In the case of the standard effective potential, the coefficients can be interpreted as the quantum corrected mass squared and couplings. We find
\begin{align}
    \label{equ:VSexpansion}
M_\mathcal{S}^2&=
        M^2+\delta M^2
        \nonumber\\
        &~~~~+\frac{3\lambda H^2}{16\pi^2}
        \left[
        \left(\nu^2-\frac{1}{4}\right)
        \left(
        \frac{2}{\varepsilon}
        +\ln\frac{4\pi\mu^2}{H^2}-\gamma
        -\psi^{(0)}\left(\frac{3}{2}-\nu\right)
        -\psi^{(0)}\left(\frac{3}{2}+\nu\right)+1
        \right)
-3
        \right]
        \nonumber\\
        &=
        M^2
        +\frac{3\lambda H^2}{16\pi^2}
        \left[
        \left(\nu^2-\frac{1}{4}\right)
        \left(
        \ln\frac{\mu^2}{H^2}
        -\psi^{(0)}\left(\frac{3}{2}-\nu\right)
        -\psi^{(0)}\left(\frac{3}{2}+\nu\right)+1
        \right)
-3
        \right],
        \nonumber\\
\lambda_\mathcal{S}&=
        \lambda+\delta\lambda-\frac{9\lambda^2}{16\pi^2}
        \left[
        \frac{2}{\varepsilon}+\ln\frac{4\pi\mu^2}{H^2}-\gamma
        -\psi^{(0)}\left(\frac{3}{2}+\nu\right)
        -\psi^{(0)}\left(\frac{3}{2}-\nu\right)+1
        \right.\nonumber\\&\left.~~~~~~~~~~~~~~~~~~~~
        +\frac{4\nu^2-1}{8\nu}\left(
        \psi^{(1)}\left(\frac{3}{2}-\nu\right)
        -\psi^{(1)}\left(\frac{3}{2}+\nu\right)
        \right)
        \right]
        \nonumber\\
        &=
        \lambda-\frac{9\lambda^2}{16\pi^2}
        \left[\ln\frac{\mu^2}{H^2}
        -\psi^{(0)}\left(\frac{3}{2}+\nu\right)
        -\psi^{(0)}\left(\frac{3}{2}-\nu\right)+1     \right.\nonumber\\&\left.~~~~~~~~~~~~~~~~~~~~
        +\frac{4\nu^2-1}{8\nu}\left(
        \psi^{(1)}\left(\frac{3}{2}-\nu\right)
        -\psi^{(1)}\left(\frac{3}{2}+\nu\right)
        \right)
        \right],
        \nonumber\\
\kappa_\mathcal{S}
        &=
        \frac{27\lambda^3}{512\pi^2\nu^3 H^2}
        \left[
        \left(12\nu^2+1\right)
        \left(
        \psi^{(1)}\left(\frac{3}{2}-\nu\right)
        -\psi^{(1)}\left(\frac{3}{2}+\nu\right)
        \right)
        \right.\nonumber\\&\left.~~~~~~~~~~~~~~~
        +\left(4\nu^2-1\right)\nu
        \left(
        \psi^{(2)}\left(\frac{3}{2}-\nu\right)
        +\psi^{(2)}\left(\frac{3}{2}+\nu\right)
        \right)
        \right],
    \end{align}
where
\begin{equation}
    \nu=\nu(0)=\sqrt{\frac{9}{4}-\frac{M^2}{H^2}}.
\end{equation}
As always, the explicit $\mu$ dependence is cancelled by the implicit $\mu$ dependence of the tree-level terms. It is instructive to consider the expansion (\ref{equ:VSexpansion}) in different mass limits. We do this in decreasing order, from large positive to large negative values of $M^2/H^2$.

\paragraph{Heavy field $M^2\gg H^2$:\\}
When $M^2>9H^2/4$, $\nu$ is imaginary, but the effective potential (\ref{equ:fullpot}) is still real. In the limit of large $M^2$, we find
\begin{align}
    \label{equ:VSheavy}
        {M}^2_\mathcal{S}
            &=M^2-\frac{3\lambda M^2}{16\pi^2}
            \left[
            \ln\frac{\mu^2}{M^2}+1
            +\frac{2H^2}{M^2}\left(\frac{7}{6}-\ln\frac{\mu^2}{M^2}\right)
            +\mathcal{O}\left(\frac{H^4}{M^4}\right)
            \right],
            \nonumber\\
        {\lambda}_\mathcal{S}
        &=\lambda
        -\frac{9\lambda^2}{16\pi^2}\left[
        \ln\frac{\mu^2}{M^2}
        +\frac{2H^2}{M^2}+\mathcal{O}\left(\frac{H^4}{M^4}\right)\right],
        \nonumber\\
        \kappa_\mathcal{S}&=\frac{27\lambda^3}{32\pi^2M^2}\left[1+\frac{2H^2}{M^2}+\mathcal{O}\left(\frac{H^4}{M^4}\right)\right].
    \end{align}
To zeroth order in $H^2/M^2$ we recover the familiar Minkowski space results. The coefficient of the sextic term $\kappa_\mathcal{S}$ is suppressed by the mass. We have also included corrections of order $H^2/M^2$, which show the leading effect of the spacetime curvature. In the case of $M_\mathcal{S}^2$ it can be interpreted as a quantum correction to the non-minimal coupling $\xi$.

\paragraph{Near conformal field $M^2\approx 2H^2$:\\}
Even though it is common to assume minimal gravitational coupling, the conformal value $\xi=1/6$ can be interesting in many cases. Therefore, we consider the case of a light, conformally coupled field, i.e. $\xi=1/6$ and $m^2\ll H^2$. 
In this case, $M^2=2H^2+m^2$ and $\nu$ is real. Expanding in powers of $m^2$, we obtain
\begin{align}
        \label{equ:VSconformal}
            {M}^2_\mathcal{S}
            &=\left(2H^2-\frac{9\lambda H^2}{16\pi^2}\right)+m^2-\frac{3\lambda m^2}{16\pi^2}
            \left[
            \ln\frac{\mu^2}{H^2}+
            2\gamma-\frac{m^2}{H^2}+
            \mathcal{O}\left(\frac{m^4}{H^4}\right)\right],
            \nonumber\\
                \lambda_\mathcal{S}
                &=\lambda-\frac{9\lambda^2}{16\pi^2}\left[
                \ln\frac{\mu^2}{H^2}+
                2\gamma
                -\frac{2m^2}{H^2}
                +
                \mathcal{O}\left(\frac{m^4}{H^4}\right)\right],
                \nonumber\\
        \kappa_\mathcal{S}&=\frac{27\lambda^3}{32\pi^2H^2}\left[2
        -\frac{12m^2}{H^2}\left(\zeta(3)-1\right)
        +\mathcal{O}\left(\frac{m^4}{H^4}\right)\right],
\end{align}
where $\zeta(z)$ is the Riemann zeta function. The conformal limit of $m^2\rightarrow 0$ is non-singular, but because the coefficient $M_\mathcal{S}^2$ of the quadratic term acquires a finite correction of order $\lambda H^2$, the loop correction breaks the conformal symmetry. The scale of the sextic term $\kappa_\mathcal{S}$ is determined by $H$. 

\paragraph{Small positive mass parameter $0<M^2\ll H^2$:\\}
The case of a light scalar field is perhaps the most interesting one for cosmology. Expanding in powers of $M^2/H^2$, we obtain
\begin{align}
\label{equ:VSlight}
 {M}^2_\mathcal{S}
            &=M^2+\frac{3\lambda H^2}{8\pi^2}\left[
            \frac{3H^2}{M^2}
            +\ln\frac{\mu^2}{H^2}+2\gamma-\frac{23}{6}
            -\frac{M^2}{2H^2}\left(\ln\frac{\mu^2}{H^2}+2\gamma+\frac{2}{27}
            \right)+\mathcal{O}\left(\frac{M^4}{H^4}\right)
            \right],
            \nonumber\\
{\lambda}_\mathcal{S}
            &=\lambda
            -\frac{9\lambda^2}{16\pi^2}
            \left[
            \frac{6H^4}{M^4}
            +\ln\frac{\mu^2}{H^2}
            +2\gamma
            +\frac{2}{27}-\frac{M^2}{2H^2}
            \left(\frac{46}{243}+\frac{16\zeta(3)}{9}\right)+\mathcal{O}\left(\frac{M^4}{H^4}\right)
            \right]  , 
            \nonumber\\
        {\kappa}_\mathcal{S}&=\frac{27\lambda^3}{32\pi^2H^2}\left[\frac{12 H^6}{M^6}+\frac{8\zeta(3)}{9}+\frac{46}{243}-\frac{M^2}{2H^2}\left(\frac{1080\zeta(3)}{729}-\frac{176}{729}\right)+\mathcal{O}\left(\frac{M^4}{H^4}\right)\right].
\end{align} 
We can see that the expressions diverge as $M^2/H^2\rightarrow 0$, indicating a breakdown of the perturbative expansion. Indeed, the relative expansion parameter in the loop expansion is $\lambda H^4/M^4$. This means that for light fields, the perturbative expansion is unlikely to converge unless $\lambda$ is very small. The coefficient of the sextic term, $\kappa_\mathcal{S}$ is large, and the coefficients of higher powers of $\bar\phi$ have even higher negative powers of $M^2$. This infrared problem means that the standard effective potential cannot be computed reliably using perturbation theory for light scalar fields.

\paragraph{Small negative mass parameter $-H^2 \ll M^2 < 0$:\\}
The case of $M^2<0$ is interesting because it is when the $\mathbb{Z}_2$ symmetry is spontaneously broken at tree level. Therefore, even at tree level and in Minkowski spacetime, the mass of the scalar particle is not given by the square root of $M^2$, which would be imaginary, but instead by the second derivative of the effective potential at its minimum.

The expansions (\ref{equ:VSlight}) are nominally valid also in this case.
However, the effective potential (\ref{equ:fullpot}) has branch point singularities at $\bar\phi=\pm (|M^2|/3\lambda)^{1/2}$
connected by a branch cut along which it has a complex value with a constant imaginary part. Therefore, the expansion around $\bar\phi=0$ is really only meaningful for $M^2>0$. Instead, with negative $M^2$, one should expand around the minimum of the potential which is at $\bar\phi=\pm(|M^2|/\lambda)^{1/2}$ at tree level. This is beyond the end of the branch cut, and therefore the expansion around that point is meaningful. However, the perturbative expansion still suffers from the same infrared problem as for positive $M^2$. 

\paragraph{Large negative mass parameter
$M^2\ll -H^2$:\\}
When $M^2\ll -H^2$, the theory approaches the Minkowski limit with a spontaneously broken $\mathbb{Z}_2$ symmetry. 
The curvature corrections are mild and suppressed by $H^2/|M^2|$.
As in the previous case, the potential has a branch cut for $|\bar\phi|\le(|M^2|/3\lambda)^{1/2}$,
and therefore the expansion around $\bar\phi=0$ in eq.~(\ref{equ:VSexp}) is not meaningful. 

\subsection{Multicomponent real scalar field}
\subsubsection{One-loop calculation}
\noindent 
It is straightforward to extend the calculation to theories with several scalar fields. As an example, we consider an $N$-component real scalar field which transforms in the fundamental representation of $O(N)$, the $N$-dimensional orthogonal group,
\begin{equation}
    \Phi=\begin{pmatrix}
       \Phi_1\\
       \vdots\\
       \Phi_N
    \end{pmatrix},
\end{equation}
where each component $\Phi_a$ with $a\in\{1,\dots,N\}$ is treated as an independent real degree of freedom. The case $N=2$ is equivalent to a single complex field. Other global symmetry groups would follow along the same lines.

For this theory, the partition function (\ref{Eq:Standard_Partition_Function}) 
becomes
\begin{equation}
\label{Eq:PartitionON_sEP}
    \mathcal{Z}_\mathcal{S}(\boldsymbol{J})=\int \mathcal{D}\Phi\,  \exp\left[{ -S_{\text{E}}[\Phi] + \int d^4x\sqrt{g}\, \left(\Phi^T\boldsymbol{J}+\boldsymbol{J}^T\Phi\right)}\right],
\end{equation}
where $T$ denotes matrix transposition and $S_{\text{E}}[\Phi]$ is the $O(N)$-invariant Euclidean action
\begin{align}
    S_{\text{E}}[\Phi]&=\int d^4x\sqrt{g}\, \left(\frac{1}{2}g^{\mu\nu} \partial_\mu \Phi^T_B\partial_\nu \Phi_B+\frac{1}{2}M_B^2(\Phi^T_B\Phi_B)+\frac{1}{4}\lambda_B(\Phi_B^T\Phi_B)^2\right)\nonumber\\
    &=\int d^4x\sqrt{g}\,\left(\frac{1}{2}Z_\phi g^{\mu\nu}\partial_\mu\Phi^T\partial_\nu\Phi
    +\frac{1}{2}\left(M^2+\delta M^2\right)(\Phi^T\Phi)
    +\frac{1}{4}\left(\lambda+\delta\lambda\right)(\Phi^T\Phi)^2
    \right),
    \label{equ:S4actionON}
\end{align}
and we demand that $\boldsymbol{J}$ transforms in the same representation as $\Phi$. 

We work in complete analogy to the singlet case to arrive at a closed form of the effective potential. We perform a field expansion about its classical expectation value,
\begin{eqnarray}
    \Phi(x)\mapsto\bar{\Phi}+\boldsymbol{\varphi}(x),
\end{eqnarray}
where we have defined two new vectors $\bar\Phi$ and $\boldsymbol{\varphi}$ analogous to eq.~(\ref{Eq:ModeEXP_sEP}). We then arrive at the quadratic action
\begin{equation}
        S_\text{E}[\Phi] = S_\text{E}[\bar\Phi] 
    + \frac{1}{2}\int d^4x\sqrt{g}\, \boldsymbol{\varphi}^T\hat{H}\boldsymbol{\varphi}+\mathcal{O}(\hbar^2),
\label{Eq:ExpandedS_sEPON}
\end{equation}
where $\hat H$ is the matrix of second-order partial derivatives at constant background field, i.e., the Hessian, with components given by
\begin{equation}       
\label{eq:Hessian}
    \hat{H}_{ab}=\left(\Delta_{\text{S}^4}+M^2+{\lambda}\bar\Phi^T\bar\Phi\right)\left[\delta_{ab}-\frac{\bar\Phi_a\bar\Phi_b}{(\bar\Phi^T\bar\Phi)}\right]+\left(\Delta_{\text{S}^4}+M^2+{3\lambda}\bar\Phi^T\bar\Phi\right)\left[\frac{\bar\Phi_a\bar\Phi_b}{(\bar\Phi^T\bar\Phi)}\right],
\end{equation}
where $a,b\in\{1,\dots,N\}$. The parametrisation above evidences the presence of two different modes with different effective masses. In particular, we find a single Higgs-like mode of mass $M^2+3\lambda\bar\Phi^T\bar\Phi$ and $(N-1)$ Goldstone-like modes with mass  $M^2+\lambda\bar\Phi^T\bar\Phi$ \cite{Shore:1979as}. We now take the logarithm of eq.~(\ref{Eq:PartitionON_sEP}) and perform a Legendre transform. This yields the $O(N)$-invariant effective potential
\begin{align}
        \mathcal{V}_{\mathcal{S}}(\bar\Phi)&=\frac{1}{2}(M^2+\delta M^2)(\bar\Phi^T\bar\Phi)+\frac{1}{4}(\lambda+\delta\lambda)(\bar\Phi^T\bar\Phi)^2\nonumber\\&\quad+\frac{1}{2\Omega_4}\ln\det\left(\frac{\Delta_{\text{S}^4} + M^2 + {3}\lambda \bar\Phi^T\bar\Phi}{H^2}\right)\nonumber\\&\quad+\frac{(N-1)}{2\Omega_4}\ln\det\left(\frac{\Delta_{\text{S}^4} + M^2 + \lambda \bar\Phi^T\bar\Phi}{H^2}\right) +\text{constant}+ \mathcal{O}(\hbar^2).   
\label{Eq:BareVsEP_sEP}
\end{align}

The regularisation and renormalisation techniques employed in the previous section easily generalise to the study of fields in non-trivial representations. As before, we begin by computing the functional determinants in eq.~(\ref{Eq:BareVsEP_sEP}). We use eq.~(\ref{Eq:F_final}) to compute the one-loop contribution for each mode, which in turn leads to two different de Sitter mass parameters, one for each of mode. In four dimensions, these read 
\begin{subequations}
    \begin{align}
    \nu_{\text{H}}&=\sqrt{\frac{9}{4}-\frac{M^2+3\lambda\bar\Phi^T\bar\Phi}{H^{2}}},\\ 
    \nu_{\text{G}}&=\sqrt{\frac{9}{4}-\frac{M^2+\lambda\bar\Phi^T\bar\Phi}{H^{2}}}.
    \end{align}
\end{subequations}
As a result, we can write the effective potential as
\begin{align}
\label{equ:ONfullpot}
    \mathcal{V}_{\mathcal{S}}(\bar\Phi)&=\frac{1}{2}(M^2+\delta M^2)(\bar\Phi^T\bar\Phi)+\frac{1}{4}(\lambda+\delta \lambda)(\bar\Phi^T\bar\Phi)^2\nonumber\\&\quad+ \mathcal{F}(\nu_{\text{H}})+(N-1) \mathcal{F}(\nu_{\text{G}})+\text{constant}+\mathcal{O}(\hbar^2),
\end{align}
where the function $\mathcal{F}$ is given by eq.~(\ref{Eq:F_final}).

\subsubsection{Power series expansions}
In line with eq.~(\ref{equ:VSexp}), we can expand the potential (\ref{equ:ONfullpot}) in powers of $\Phi$,
\begin{equation}
    \label{equ:VSexpON}
    \mathcal{V}_\mathcal{S}(\bar\Phi)=\frac{1}{2} M^2_\mathcal{S}(\bar\Phi^T\bar\Phi)+\frac{1}{4}\lambda_\mathcal{S}(\bar\Phi^T\bar\Phi)^2+\frac{1}{6}\kappa_\mathcal{S}(\bar\Phi^T\bar\Phi)^3+\mathcal{O}\left((\bar\Phi^T\bar\Phi)^4\right),
\end{equation}
and calculate the coefficients for the same cases as in section~\ref{sec:powerseries}.
The qualitative behaviour is the same as for a single real field, and therefore we only quote the expansions.
\paragraph{Heavy field $|M^2|\gg H^2$:}
\begin{align}
    \label{equ:VSheavyON}
        {M}^2_\mathcal{S}
            &=M^2-\frac{(N+2)\lambda M^2}{16\pi^2}
            \left[
            \ln\frac{\mu^2}{M^2}+1
            +\frac{2H^2}{M^2}\left(\frac{7}{6}-\ln\frac{\mu^2}{M^2}\right)
            +\mathcal{O}\left(\frac{H^4}{M^4}\right)
            \right],
            \nonumber\\
        {\lambda}_\mathcal{S}
        &=\lambda
        -\frac{(N+8)\lambda^2}{16\pi^2}\left[
        \ln\frac{\mu^2}{M^2}
        +\frac{2H^2}{M^2}+\mathcal{O}\left(\frac{H^4}{M^4}\right)\right],
        \nonumber\\
        \kappa_\mathcal{S}&=\frac{(N+26)\lambda^3}{32\pi^2M^2}\left[1+\frac{2H^2}{M^2}+\mathcal{O}\left(\frac{H^4}{M^4}\right)\right].
    \end{align}
\paragraph{Near conformal field $M^2\approx2H^2$:}
\begin{align}
        \label{equ:VSconformalON}
            {M}^2_\mathcal{S}
            &=\left(2H^2-\frac{3(N+2)\lambda H^2}{16\pi^2}\right)+m^2-\frac{(N+2)\lambda m^2}{16\pi^2}
            \left[
            \ln\frac{\mu^2}{H^2}+
            2\gamma-\frac{m^2}{H^2}+
            \mathcal{O}\left(\frac{m^4}{H^4}\right)\right],
            \nonumber\\
                \lambda_\mathcal{S}
                &=\lambda-\frac{(N+8)\lambda^2}{16\pi^2}\left[
                \ln\frac{\mu^2}{H^2}+
                2\gamma
                -\frac{2m^2}{H^2}
                +
                \mathcal{O}\left(\frac{m^4}{H^4}\right)\right],
                \nonumber\\
        \kappa_\mathcal{S}&=\frac{(N+26)\lambda^3}{32\pi^2H^2}\left[2
        -\frac{12m^2}{H^2}\left(\zeta(3)-1\right)
        +\mathcal{O}\left(\frac{m^4}{H^4}\right)\right],
\end{align}

\paragraph{Light field $|M^2|\ll H^2$:}
\begin{align}
\label{equ:VSlightON}
 {M}^2_\mathcal{S}
            &=M^2+\frac{(N+2)\lambda H^2}{8\pi^2}\left[
            \frac{3H^2}{M^2}
            +\ln\frac{\mu^2}{H^2}+2\gamma-\frac{23}{6}\right.\nonumber\\&\left.
            ~~~~~~~~~~~~~~~~~~~~~~~~~~~~~~~
            -\frac{M^2}{2H^2}\left(\ln\frac{\mu^2}{H^2}+2\gamma+\frac{2}{27}
            \right)
            +\mathcal{O}\left(\frac{M^4}{H^4}\right)
            \right],
            \nonumber\\
{\lambda}_\mathcal{S}
            &=\lambda
            -\frac{(N+8)\lambda^2}{16\pi^2}
            \left[
            \frac{6H^4}{M^4}
            +\ln\frac{\mu^2}{H^2}
            +2\gamma
            +\frac{2}{27}-\frac{M^2}{2H^2}
            \left(\frac{46}{243}+\frac{16\zeta(3)}{9}\right)+\mathcal{O}\left(\frac{M^4}{H^4}\right)
            \right]  , 
            \nonumber\\
        {\kappa}_\mathcal{S}&=\frac{(N+26)\lambda^3}{32\pi^2H^2}\left[\frac{12 H^6}{M^6}+\frac{8\zeta(3)}{9}+\frac{46}{243}-\frac{M^2}{2H^2}\left(\frac{1080\zeta(3)}{729}-\frac{176}{729}\right)+\mathcal{O}\left(\frac{M^4}{H^4}\right)\right].
\end{align}    
       
\noindent We see that in the case $N=1$, we recover the single scalar field results, as expected.

\section{Constraint effective potential\label{Sec:cEP}}
\subsection{Definition
\label{Sec:EffPots_cEP}}
The constraint effective potential~\cite{ORaifeartaigh:1986axd} is defined by inserting a delta functional in the partition function. This gives the constraint partition function,
\begin{equation}
\label{Eq:constraint def}
    \mathcal{Z}_{\mathcal{C}}(\bar\phi)=\int \mathcal{D}\phi\, \delta\!\left(\bar\phi-\frac{1}{\Omega_4}\int d^4x\sqrt{g}\,\phi(x)\right)\,e^{-S[\phi]}.
\end{equation}
The delta function here plays the role of a holonomic constraint, so the path integral is only over configurations in which the average value of the field over the manifold is equal to $\bar\phi$. The constraint effective potential is then defined directly as
\begin{equation}
\label{equ:VCdef}
    \mathcal{V}_\mathcal{C}(\bar\phi)=-\frac{1}{\Omega_4}\,\ln \mathcal{Z}_\mathcal{C}(\bar\phi).
\end{equation}
Concretely, it is the effective potential of the averaged field $\bar\phi$ obtained by integrating out all inhomogeneous fluctuations. This is analogous to the second definition (\ref{equ:F2}) of the Helmholtz free energy, in which the particle number remains fixed.

It follows directly from the definition (\ref{equ:VCdef}) that the probability distribution of $\bar\phi$, the field averaged over the whole space, is given by
\begin{equation}
\label{equ:pphi}
    p(\bar\phi)\propto \exp\left(
    -\frac{8\pi^2\mathcal{V}_\mathcal{C}(\bar\phi)}{3H^4}
    \right).
\end{equation}
Therefore the minimum of $\mathcal{V}_\mathcal{C}$ gives the most likely value of $\bar\phi$. In contrast, the minimum of the standard effective potential $\mathcal{V}_\mathcal{S}$ gives the expectation value of the field, $\langle\phi\rangle$,  which is related to eq.~(\ref{equ:pphi}) through
\begin{equation}
\langle \phi\rangle = \int d\bar\phi\,\bar\phi\,  p(\bar\phi).
\end{equation}
In the infinite-volume limit, the probability distribution (\ref{equ:pphi}) becomes infinitesimally narrow, and therefore these two become equivalent, but in a finite volume they are different.

It is important to note that our spacetime has a finite volume only because of the analytic continuation to Euclidean signature (\ref{equ:spheremetric}), as the original de Sitter spacetime has infinite volume. It is not clear if the constraint effective potential, or even the averaged field $\bar\phi$ can be expressed directly in terms of the Lorentzian de Sitter degrees of freedom. Therefore, the analytic continuation to Euclidean signature should be considered to be an integral step in the definition of the constraint effective potential.

A more complete relation between the two potentials is that the function (\ref{equ:WSJ}) used to define $\mathcal{V}_\mathcal{S}$ can be expressed as
\begin{equation}
\label{equ:VCtoWSJ}
    \mathcal{W}(J)
    =-\frac{3H^4}{8\pi^2}\ln\left\{\int d\bar\phi\, \exp\left[
    -\frac{8\pi^2}{3H^4}\left(
    \mathcal{V}_\mathcal{C}(\bar\phi)-J\bar\phi
    \right)
    \right]\right\}.
\end{equation}

\subsection{Real scalar singlet}
\subsubsection{One-loop calculation}
To compute the constraint partition function (\ref{Eq:constraint def}) for a single real scalar field, we express the delta function in terms of its Fourier representation, obtaining
\begin{equation}
    \mathcal{Z}_\mathcal{C}(\bar\phi)=\int \mathcal{D}\phi
    \int_{-\infty}^\infty dB\, \exp\left[-S_\text{E}[\phi]-\frac{iB}{\Omega_4}\int d^4x\sqrt{g}\,(\phi(x)-\bar\phi)\right].
\end{equation}
To evaluate this at one loop, we again write the field in terms of the background $\bar\phi$ and fluctuations $\varphi$,
\begin{equation}
    \phi(x)=\bar\phi+\varphi(x),
\end{equation}
and arrive at the quadratic action
\begin{equation}
    \mathcal{Z}_\mathcal{C}(\bar\phi)=e^{-S_\text{E}[\bar\phi]}\int \mathcal{D}\varphi\int_{-\infty}^\infty dB\,\exp\left[
    - \frac{1}{2}\int d^4x\sqrt{g}\, \left(\varphi\hat{O}(\bar\phi)\varphi+\frac{2iB\varphi}{\Omega_4}\right)+\mathcal{O}(\hbar^2)\right],
\end{equation}
where $\hat{O}(\bar\phi)$ is the same operator as in eq.~(\ref{Eq:EKG}). We complete the square as
\begin{equation}
    \tilde{\varphi}(x)=\varphi(x)+\frac{iB}{\Omega_4\omega^2_0},
\end{equation}
where $\omega^2_0=V''(\bar\phi)=M^2+3\lambda\bar\phi^2$ is the eigenvalue of $\hat{O}(\bar\phi)$ corresponding to the constant eigenfunction $\varphi_0(x)=1$ (see eq.~(\ref{Eq:SpectrumLaplacian})). As a result, the partition function factorises into
\begin{align}
    \mathcal{Z}_\mathcal{C}(\bar\phi)&=e^{-S_\text{E}[\bar\phi]}\times\left(\int \mathcal{D}\tilde\varphi\, \exp\left[
    - \frac{1}{2}\int d^4x\sqrt{g}\, \tilde\varphi\hat{O}(\bar\phi)\tilde\varphi\right]\right)\times
    \left(\int_{-\infty}^\infty dB\,\exp\left[-\frac{B^2}{\Omega_4\omega^2_0}\right]\right)
    \nonumber\\
    &=e^{-S_\text{E}[\bar\phi]}\times{\frac{1}{\sqrt{\det\left(\hat{O}(\bar\phi)/{H^2}\right)}}}\times\sqrt{{\pi }{{\Omega_4 \omega_0^2}}}\times\text{constant}\times\mathcal{O}(\hbar^2).
\end{align}
Equation (\ref{equ:VCdef}) then gives us the constraint effective potential as
\begin{align}
\label{equ:VCresult}
     \mathcal{V}_{\mathcal{C}}(\bar\phi)&=
     V(\bar\phi)+\delta V(\bar\phi)
     +\frac{1}{2\Omega_4}
     \left[\ln\det\left({\frac{\hat{O}(\bar\phi)}{H^2}}\right)-\ln\frac{\omega_0^2}{H^2}\right]+\text{constant}+\mathcal{O}(\hbar^2).
\end{align}
Up to irrelevant additive constants, the second term inside the square brackets removes the $n=0$ contribution from the series~(\ref{Eq:I_1}), so that the remaining sum starts at $n=1$. This shows that the difference between eqs.~(\ref{equ:VCresult}) and~(\ref{Eq:BareVsEP_sEP}) arises solely from the homogeneous mode. Since this long-wavelength mode is ultraviolet finite, the constrained effective potential is also finite. As a result, we may relate the two potentials through
\begin{equation}
\label{equ:VCvVs}
    \mathcal{V}_{\mathcal{C}}(\bar\phi)
    =\mathcal{V}_{\mathcal{S}}(\bar\phi)-\frac{1}{2\Omega_4}\ln\left(1+\frac{3\lambda\bar\phi^2}{M^2}\right)+
    \text{constant}+\mathcal{O}(\hbar^2).
\end{equation}
Therefore, using eq.~(\ref{equ:fullpot}), the complete one-loop expression for the constraint effective potential is
\begin{align}
\label{equ:fullCpot}
\mathcal{V}_\mathcal{C}(\bar\phi)
    &=\frac{1}{2}M^2\bar\phi^2+\frac{1}{4}\lambda\bar\phi^4
    \nonumber\\
    &
    \quad-\frac{H^4}{64\pi^2}
        \left\{ \left(\nu^4{(\bar\phi)}-\frac{\nu^2{(\bar\phi)}}{2}\right)
        \left(\ln\frac{\mu^2}{H^2}
        +1\right)
        \right.
        \nonumber\\
        &\quad\quad+6\nu^2{(\bar\phi)}-\left(4\nu^2{(\bar\phi)}-\nu{(\bar\phi)}\right)\left[\ln\!\Gamma\left(\frac{3}{2}+{\nu{(\bar\phi)}}\right)-\ln\!\Gamma\left(\frac{3}{2}-{\nu{(\bar\phi)}}\right)\right]
        \nonumber\\
        &\quad\quad
        +\left(12\nu^2{(\bar\phi)}-1\right)\left[\psi^{(-2)}\left(\frac{3}{2}+{\nu{(\bar\phi)}}\right)+\psi^{(-2)}\left(\frac{3}{2}-{\nu{(\bar\phi)}}\right)\right]
        \nonumber\\
         &\quad\quad-24\nu{(\bar\phi)}\left[\psi^{(-3)}\left(\frac{3}{2}+{\nu{(\bar\phi)}}\right)-\psi^{(-3)}\left(\frac{3}{2}-{\nu{(\bar\phi)}}\right)\right]
         \nonumber\\
         &\left.\quad\quad+24\left[\psi^{(-4)}\left(\frac{3}{2}+{\nu{(\bar\phi)}}\right)+\psi^{(-4)}\left(\frac{3}{2}-{\nu{(\bar\phi)}}\right)\right]\right.
         \nonumber\\
         &\left.\quad\quad+12\ln\left(\frac{9}{4}-\nu(\bar\phi)^2\right)
         \right\}+\text{constant}+\mathcal{O}(\hbar^2),
\end{align}
where, again, $\nu(\bar\phi)$ is given by eq.~(\ref{equ:nuofphi}).

\subsubsection{Power series expansions}
As in eq.~(\ref{equ:VSexp}), we expand the constrained effective potential as a power series in the field,
\begin{equation}
    \label{equ:VCexp}
    \mathcal{V}_\mathcal{C}(\bar\phi)=\frac{1}{2} M^2_\mathcal{C}\bar\phi^2+\frac{1}{4}\lambda_\mathcal{C}\bar\phi^4+\frac{1}{6}\kappa_\mathcal{C}\bar\phi^6+\mathcal{O}(\bar\phi^8).
\end{equation}
We use the symbols $M^2_\mathcal{C}$, $\lambda_\mathcal{C}$, and $\kappa_\mathcal{C}$ for the coefficients, but this is only for the sake of familiarity. In this case these coefficients do not have the same interpretation as quantum corrected mass parameters or couplings. Expanding eq.~(\ref{equ:VCvVs}) in powers of $\bar\phi$, we find
\begin{align}
\label{equ:CvsS}
M_\mathcal{C}^2
        &=M_\mathcal{S}^2
        -\frac{9\lambda H^4}{8\pi^2 M^2}+\mathcal{O}(\lambda^2),
        \nonumber\\
\lambda_\mathcal{C}&=\lambda_\mathcal{S}+\frac{27\lambda^2H^4}{8\pi^2M^4}+\mathcal{O}(\lambda^3),
        \nonumber\\
\kappa_\mathcal{C}
        &=\kappa_\mathcal{S}-\frac{81\lambda^3 H^4}{8\pi^2 M^6}+\mathcal{O}(\lambda^4),
\end{align}
where $M^2_\mathcal{S}$, $\lambda_\mathcal{S}$ and $\kappa_\mathcal{S}$ are given in eq.~(\ref{equ:VSexpansion}). 

It is now instructive to compare the behaviour of these coefficients in the mass ranges that were discussed in section~\ref{sec:powerseries}.

\paragraph{Heavy fields $M^2\gg H^2:$\\}
In this case, we obtain
\begin{align}
\label{equ:VCheavy}
{M}^2_\mathcal{C}
            &=M^2-\frac{3\lambda M^2}{16\pi^2}
            \left[
            \ln\frac{\mu^2}{M^2}+1
            +\frac{2H^2}{M^2}\left(\frac{7}{6}-\ln\frac{\mu^2}{M^2}\right)
            +O\left(\frac{H^4}{M^4}\right)
            \right],\nonumber\\
{\lambda}_\mathcal{C}
        &=\lambda
        -\frac{9\lambda^2}{16\pi^2}\left[
        \ln\frac{\mu^2}{M^2}
        +\frac{2H^2}{M^2}
        +\mathcal{O}\left(\frac{H^4}{M^4}\right)\right],
        \nonumber\\
\kappa_\mathcal{C}&=\frac{27\lambda^3}{32\pi^2M^2}\left[1+
        \frac{2H^2}{M^2}+
        \mathcal{O}\left(\frac{H^4}{M^4}\right)\right].
\end{align}
These are the same as eq.~(\ref{equ:VSheavy}) for the standard effective potential, because the difference only appears at order $\mathcal{O}(H^4/M^4)$. This is to be expected because the two effective potentials are equal in the infinite-volume limits, and in this case the de Sitter radius is much larger than the inverse mass.

\paragraph{Near conformal field $M^2\approx 2H^2$:\\}
For the conformally coupled field, i.e., $\xi=1/6$ and $m^2\ll H^2$, we find
\begin{align}
        \label{equ:VCconformal}
            {M}^2_\mathcal{C}
            &=\left(2H^2-\frac{9\lambda H^2}{8\pi^2}\right)+m^2-\frac{3\lambda m^2}{16\pi^2}
            \left[
            \ln\frac{\mu^2}{H^2}+
            2\gamma
            -\frac{3}{2}
            -\frac{m^2}{4H^2}
            +\mathcal{O}\left(\frac{m^4}{H^4}\right)\right],
            \nonumber\\
\lambda_\mathcal{C}
                &=\lambda-\frac{9\lambda^2}{16\pi^2}\left[
                \ln\frac{\mu^2}{H^2}+
                2\gamma
                -\frac{3}{2}
                -\frac{m^2}{2H^2}+
                \mathcal{O}\left(\frac{m^4}{H^4}\right)\right],
                \nonumber\\
        \kappa_\mathcal{C}&=\frac{27\lambda^3}{32\pi^2H^2}\left[\frac{1}{2}-\frac{3m^2}{4H^2}
        \left(16\zeta(3)-19\right)+        
        \mathcal{O}\left(\frac{m^4}{H^4}\right)\right].
\end{align}
At the detailed level, these parameters differ from eq.~(\ref{equ:VSconformal}),
but quantitatively the behaviour is the same.

\paragraph{Small positive mass parameter
$0<M^2\ll H^2$:\\}
For light fields, $M^2\ll H^2$, we find
\begin{align}
\label{equ:VClight}
 {M}^2_\mathcal{C}
            &=M^2+\frac{3\lambda H^2}{8\pi^2}\left[\ln\frac{\mu^2}{H^2}+2\gamma-\frac{23}{6}
            -\frac{M^2}{2H^2}\left(\ln\frac{\mu^2}{H^2}+2\gamma+\frac{2}{27}
            \right)+\mathcal{O}\left(\frac{M^4}{H^4}\right)
            \right],
            \nonumber\\
{\lambda}_\mathcal{C}
            &=\lambda
            -\frac{9\lambda^2}{16\pi^2}
            \left[\ln\frac{\mu^2}{H^2}
            +2\gamma
            +\frac{2}{27}-\frac{M^2}{2H^2}
            \left(\frac{46}{243}+\frac{16\zeta(3)}{9}\right)+\mathcal{O}\left(\frac{M^4}{H^4}\right)
            \right]  , 
            \nonumber\\
{\kappa}_\mathcal{C}&=
            \frac{27\lambda^3}{32\pi^2H^2}\left[\frac{8\zeta(3)}{9}+\frac{46}{243}-\frac{M^2}{2H^2}\left(\frac{1080\zeta(3)}{729}-\frac{176}{729}\right)+\mathcal{O}\left(\frac{M^4}{H^4}\right)\right].
\end{align}   
In this case, which is often the relevant one for cosmology, we see a significant difference compared to the coefficients (\ref{equ:VSlight}) of the standard effective potential. The terms in eq.~(\ref{equ:CvsS}) cancel the negative powers of $M^2$ exactly, and we are left with expressions that remain finite and small for arbitrarily small $M^2/H^2$. This means that the constraint effective potential does not suffer from the infrared problem, and that it can computed using perturbation theory as long as $\lambda\ll 1$.

\paragraph{Small negative mass parameter $-H^2\ll M^2<0$:\\}
The same expansion (\ref{equ:VClight}) is also valid for small negative mass parameter. Furthermore, the subtraction in eq.~(\ref{equ:VCvVs}) cancels the branch cut singularity. Therefore the constraint effective potential is real and well defined for all field values, unlike the standard effective potential. Because it does not suffer from the infrared problem either and can therefore be computed using perturbation theory, it can be a useful tool for studying spontaneous symmetry breaking in de Sitter, as long as one remembers that its physical interpretation is different from the standard effective potential.

\paragraph{Large negative mass  parameter $M^2\ll -H^2$:\\}
Just like for large positive $M^2$, this case is identical to the standard effective potential. Therefore the discussion in section~\ref{sec:powerseries} applies here, too.

\subsection{Multicomponent real scalar field}
\subsubsection{One-loop calculation}
The constraint partition function for a scalar field $N$-plet is given by
\begin{equation}
    \mathcal{Z}_\mathcal{C}(\bar\Phi)=\int \mathcal{D}\Phi\int d\boldsymbol{B}\,  \exp\left[-S[\Phi]-\frac{i\boldsymbol{B}^T}{\Omega_4}\int d^4x\sqrt{g}\,(\Phi(x)-\bar\Phi)\right],
\end{equation}
where in this case $\boldsymbol{B}$ is a $N$-dimensional vector that imposes the constraint (\ref{Eq:constraint def}) individually for each mode in $\Phi$. Upon, expanding about the background as in eq.~(\ref{Eq:ModeEXP_sEP}), we find the one-loop action
\begin{equation}
    \mathcal{Z}_\mathcal{C}(\bar\Phi)=e^{-S_\text{E}[\bar\Phi]}\int \mathcal{D}\boldsymbol{\varphi}\int_{-\infty}^\infty d\boldsymbol{B}\,\exp\left[
    - \frac{1}{2}\int d^4x\sqrt{g}\, \left(\boldsymbol{\varphi}^T\hat{H}\boldsymbol{\varphi}+\frac{2i\boldsymbol{B}^T\boldsymbol{\varphi}}{\Omega_4}\right)+\mathcal{O}(\hbar^2)\right],
\end{equation}
where $\hat{H}$ is the Hessian matrix in (\ref{eq:Hessian}). This time, we complete the square as
\begin{equation}
    \tilde{\varphi}_a(x)=\varphi_a(x)+\frac{iB_a}{\Omega_4\omega^2_{a,0}},
\end{equation}
with
\begin{align}
    \omega_{a,0}^2=\begin{cases}
        M^2+3\lambda(\bar\Phi^T\bar\Phi)&a=1\\
        M^2+\lambda(\bar\Phi^T\bar\Phi) &a>1
    \end{cases},
\end{align}
where $\omega_{a,0}^2$ are the eigenvalue of the Hessian matrix corresponding to the constant Higgs-like $(a=1)$ and Goldstone-like $(a>1)$ eigenfunctions, respectively. As a result, the constraint effective potential takes the form
\begin{align}
\label{equ:VCresultON}
     \mathcal{V}_{\mathcal{C}}(\bar\Phi)&=\frac{1}{2}(M^2+\delta M^2)(\bar\Phi^T\bar\Phi)+\frac{1}{4}(\lambda+\delta\lambda)(\bar\Phi^T\bar\Phi)^2\nonumber\\
     &\quad+\frac{1}{2\Omega_4}
     \left[\ln\det\left({\frac{\Delta_{\text{S}^4}+M^2+3\lambda\bar\Phi^T\bar\Phi}{H^2}}\right)-\ln\left(\frac{M^2+3\lambda\bar\Phi^T\bar\Phi}{H^2}\right)\right]\nonumber\\&\quad+\frac{(N-1)}{2\Omega_4}
     \left[\ln\det\left({\frac{\Delta_{\text{S}^4}+M^2+\lambda\bar\Phi^T\bar\Phi}{H^2}}\right)-\ln\left(\frac{ M^2+\lambda\bar\Phi^T\bar\Phi}{H^2}\right)\right]\nonumber\\
     &\quad+\text{constant}+\mathcal{O}(\hbar^2).   
\end{align}
Alternatively, we can write
\begin{equation}
    \mathcal{V}_{\mathcal{C}}(\bar\Phi)=\mathcal{V}_{\mathcal{S}}(\bar\Phi)-\frac{1}{2\Omega_4} \ln\left(1+\frac{3\lambda\bar\Phi^T\bar\Phi}{M^2}\right)-\frac{(N-1)}{2\Omega_4}\ln\left(1+\frac{\lambda\bar\Phi^T\bar\Phi}{M^2}\right)+\text{constant}+\mathcal{O}(\hbar^2).
\end{equation}

\subsubsection{Power series expansions}
Again, we expand the potential in powers of the field $\bar\Phi$,
\begin{equation}
    \label{equ:VCexpON}
    \mathcal{V}_\mathcal{C}(\bar\Phi)=\frac{1}{2} M^2_\mathcal{C}(\bar\Phi^T\bar\Phi)+\frac{1}{4}\lambda_\mathcal{C}(\bar\Phi^T\bar\Phi)^2+\frac{1}{6}\kappa_\mathcal{C}(\bar\Phi^T\bar\Phi)^3+\mathcal{O}\left((\bar\Phi^T\bar\Phi)^4\right),
\end{equation}
and quote the coefficients in the different mass ranges.
\paragraph{Heavy field $|M^2|\gg H^2$:}
\begin{align}
{M}^2_\mathcal{C}
            &=M^2-\frac{(N+2)\lambda M^2}{16\pi^2}
            \left[
            \ln\frac{\mu^2}{M^2}+1
            +\frac{2H^2}{M^2}\left(\frac{7}{6}-\ln\frac{\mu^2}{M^2}\right)
            +O\left(\frac{H^4}{M^4}\right)
            \right],\nonumber\\
{\lambda}_\mathcal{C}
        &=\lambda
        -\frac{(N+8)\lambda^2}{16\pi^2}\left[
        \ln\frac{\mu^2}{M^2}
        +\frac{2H^2}{M^2}
        +\mathcal{O}\left(\frac{H^4}{M^4}\right)\right],
        \nonumber\\
\kappa_\mathcal{C}&=\frac{(N+26)\lambda^3}{32\pi^2M^2}\left[1+
        \frac{2H^2}{M^2}+
        \mathcal{O}\left(\frac{H^4}{M^4}\right)\right].
\end{align}
\paragraph{Near conformal field $M^2\approx 2H^2$:}
\begin{align}
            {M}^2_\mathcal{C}
            &=\left(2H^2-\frac{3(N+2)\lambda H^2}{8\pi^2}\right)+m^2
            \nonumber\\&
            ~~~~-\frac{(N+2)\lambda m^2}{16\pi^2}
            \left[
            \ln\frac{\mu^2}{H^2}+
            2\gamma
            -\frac{3}{2}
            -\frac{m^2}{4H^2}
            +\mathcal{O}\left(\frac{m^4}{H^4}\right)\right],
            \nonumber\\
\lambda_\mathcal{C}
                &=\lambda-\frac{(N+8)\lambda^2}{16\pi^2}\left[
                \ln\frac{\mu^2}{H^2}+
                2\gamma
                -\frac{3}{2}
                -\frac{m^2}{2H^2}+
                \mathcal{O}\left(\frac{m^4}{H^4}\right)\right],
                \nonumber\\
        \kappa_\mathcal{C}&=\frac{(N+26)\lambda^3}{32\pi^2H^2}\left[\frac{1}{2}-\frac{3m^2}{4H^2}
        \left(16\zeta(3)-19\right)+        
        \mathcal{O}\left(\frac{m^4}{H^4}\right)\right].
\end{align}
\paragraph{Light field $|M^2|\ll 2H^2$:}
\begin{align}
 {M}^2_\mathcal{C}
            &=M^2+\frac{(N+2)\lambda H^2}{8\pi^2}\left[\ln\frac{\mu^2}{H^2}+2\gamma-\frac{23}{6}
            -\frac{M^2}{2H^2}\left(\ln\frac{\mu^2}{H^2}+2\gamma+\frac{2}{27}
            \right)+\mathcal{O}\left(\frac{M^4}{H^4}\right)
            \right],
            \nonumber\\
{\lambda}_\mathcal{C}
            &=\lambda
            -\frac{(N+8)\lambda^2}{16\pi^2}
            \left[\ln\frac{\mu^2}{H^2}
            +2\gamma
            +\frac{2}{27}-\frac{M^2}{2H^2}
            \left(\frac{46}{243}+\frac{16\zeta(3)}{9}\right)+\mathcal{O}\left(\frac{M^4}{H^4}\right)
            \right]  , 
            \nonumber\\
{\kappa}_\mathcal{C}&=
            \frac{(N+26)\lambda^3}{32\pi^2H^2}\left[\frac{8\zeta(3)}{9}+\frac{46}{243}-\frac{M^2}{2H^2}\left(\frac{1080\zeta(3)}{729}-\frac{176}{729}\right)+\mathcal{O}\left(\frac{M^4}{H^4}\right)\right].
\end{align}   

\section{Discussion\label{Sec:Discussion}}
\subsection{Infrared problem}
It is clear from our results for light fields in eqs.~(\ref{equ:VSlight}) and (\ref{equ:VClight}) that the constraint effective potential does not suffer from the same infrared problem as the standard effective potential. Although our explicit calculation was only done to one loop order, this finding would appear to be valid to all orders. For the standard effective potential, adding a tadpole loop to a propagator line in the loop expansion gives a factor that is parametrically of order
\begin{equation}
\label{equ:IRdiv}
    \frac{\lambda H^4}{(\omega_0^2)^2}\sim \lambda\frac{H^4}{M^4},
\end{equation}
where $\omega_0^2=V''(\bar\phi)$ is the lowest eigenvalue (\ref{equ:eveq}) of the Hessian, $\hat{O}(\bar\phi)$.
For light fields, $M\ll \lambda^{1/4}H$, eq.(\ref{equ:IRdiv}) becomes larger than one, and perturbation theory is therefore not applicable. In the case of the constraint effective potential, the constraint removes the $n=0$ term from the sum (\ref{Eq:I_1}), so the contribution from the tadpole is
\begin{equation}
    \frac{\lambda H^4}{(\omega_1^2)^2}\sim \lambda,
\end{equation}
with $\omega_1^2\approx 4H^2$ for light fields. 
More generally, because the $n=0$ is absent, no contributions of the form (\ref{equ:IRdiv}) can arise from other diagrams either. Therefore we expect that the constraint effective potential is well approximated by the one-loop result, as long as $\lambda\ll1$.

The reason for the difference is that, just like the two free energies (\ref{equ:F1}) and (\ref{equ:F2}), the standard and constraint effective potential describe different aspects of the same physical system. 
For light but not massless fields, the probability distribution (\ref{equ:pphi}) and the constraint effective potential are well defined, finite, and perturbatively calculable. The infrared problem is confined in the integral (\ref{equ:VCtoWSJ}), which cannot be computed as a power series in the coupling $\lambda$ if the field is light.

\subsection{Stochastic theory}
When interpreting the constraint effective potential through eq.~(\ref{equ:pphi}), one should be careful about what quantity it is the probability distribution of. The variable $\bar\phi$ is the field averaged over the four-sphere obtained as a Wick rotation of de Sitter spacetime to Euclidean signature. Therefore, it has no direct link to physical observables, which are defined in the original de Sitter spacetime. It is, nevertheless, interesting to observe that the stochastic approach~\cite{Starobinsky:1986fx,Starobinsky:1994bd} appears to provide an indirect link, as we will now demonstrate.

In the stochastic approach, the dynamics of a light scalar field on large distances in a de Sitter spacetime is described by the stochastic Langevin equation
\begin{equation}
\label{equ:stochastic}
    \frac{d\phi}{dt}+\frac{\mathcal{V}'(\phi)}{3H}=\xi,
\end{equation}
where $\xi$ is a white noise term that satisfies 
\begin{equation}
    \langle\xi(t)\xi(t')\rangle
    =\frac{H^3}{4\pi^2}\delta\!\left(t-t'\right),
\end{equation}
and $\mathcal{V}$ is the potential of the scalar field.

One obvious question that arises from eq.~(\ref{equ:stochastic}) is what should be used as the potential $\mathcal{V}(\phi)$. 
The original derivation of the stochastic theory by Starobinsky and Yokoyama~\cite{Starobinsky:1986fx,Starobinsky:1994bd} treated the quantum fields at linear order, and therefore they obtained a Langevin equation with the tree-level potential. Taken literally, that would mean that the potential $\mathcal{V}(\phi)$ is the bare potential $V_B(\phi_B)$ of the quantum field theory. However, this cannot be correct because the latter is ultraviolet divergent. In the quantum field theory, these divergences, expressed using the counterterms $\delta M^2$ and $\delta\lambda$, cancel the ultraviolet divergences arising from loop corrections, so that the physical observables are finite. In the stochastic theory (\ref{equ:stochastic}), there are no ultraviolet divergent loops, and therefore nothing that could cancel them. And it cannot be the renormalised tree-level potential $V(\phi)$ either, because its coefficients depend on the renormalisation scheme and scale. Therefore it should be some kind of an effective potential, but which one?

Whatever the potential $\mathcal{V}(\phi)$ is, the Fokker-Planck equation corresponding to eq.~(\ref{equ:stochastic}) can be solved to find that in equilibrium, the field $\phi$ has the one-point probability distribution~\cite{Starobinsky:1986fx,Starobinsky:1994bd}
\begin{equation}
        p_{\rm stoch}(\phi)\propto \exp\left(
    -\frac{8\pi^2\mathcal{V}(\phi)}{3H^4}
    \right).
\end{equation}
If we identify the stochastic field variable $\phi$ with the averaged quantum field $\bar\phi$, comparison with eq.~(\ref{equ:pphi}) shows that the potential in eq.~(\ref{equ:stochastic}) should be chosen to be precisely the constraint effective potential, $\mathcal{V}=\mathcal{V}_\mathcal{C}$.
However, the caveat is that the derivation of eq.~(\ref{equ:stochastic}) in ref.~\cite{Starobinsky:1994bd} is not precise about the definition of the stochastic variable $\phi$, and therefore it is not clear that it is physically accurate to assume $\phi=\bar\phi$.

Because of this, it is useful to go beyond the field variable and consider actual physical observables. In ref.~\cite{Camargo-Molina:2022paw}, the stochastic theory (\ref{equ:stochastic}) was used to compute the lifetime of a metastable vacuum. This can be done precisely using a spectral expansion or, alternatively, using the saddle-point approximation. The authors found that the latter result agrees exactly with the result of the saddle point approximation in quantum field theory, based on Hawking-Moss instantons~\cite{Hawking:1981fz}, if the potential in eq.~(\ref{equ:stochastic}) is chosen to be the constraint effective potential, i.e., if $\mathcal{V}=\mathcal{V}_\mathcal{C}$.

We can also compare correlation functions. Because the stochastic theory is only valid for light fields, $M\ll H$, we have to limit our comparison to that case. We will compute the two-point correlation function of the scalar field $\phi$ at one loop in perturbation theory, in both quantum field theory and the stochastic theory. Because of the infrared problem, the perturbative expansion is only valid when $\lambda\ll M^4/H^4$, but it still provides a non-trivial check.

In quantum field theory, the one-loop correction to the correlation function is given by the tadpole diagram, which amounts to a simple mass correction. Therefore, the one-loop correlation function is given by the tree-level propagator with the mass parameter replaced by the quantum corrected mass parameter. Conveniently for us, the latter is precisely the coefficient $M_\mathcal{S}^2$ given in eq.~(\ref{equ:VSexpansion}). Using the tree-level propagator from ref.~\cite{Bunch:1978yq}, the one-loop time-ordered correlator for a light field at asymptotically long distances is therefore
\begin{align}
\label{equ:qftcorr}
    \langle \mathcal{T}\phi(x)\phi(y)\rangle_{\rm QFT}
    &\sim
    \frac{H^2}{16\pi^2}\frac{\Gamma(3/2-\nu_\mathcal{S})\Gamma(2\nu_\mathcal{S})}{\Gamma(1/2+\nu_\mathcal{S})}
    \left(\frac{z}{2}\right)^{-\left(\frac{3}{2}-\nu_\mathcal{S}\right)}
    +\mathcal{O}(\lambda^2)
    \nonumber\\
    &= \left(\frac{3H^4}{8\pi^2M_\mathcal{S}^2}
   +\mathcal{O}(H^2)\right)
    \left(\frac{z}{2}\right)^{-\frac{M_\mathcal{S}^2
    }{3H^2}+\mathcal{O}\left(\frac{M_\mathcal{S}^4}{H^4}\right)}
    +\mathcal{O}(\lambda^2),
\end{align}
where 
\begin{equation}
    \nu_\mathcal{S}=\sqrt{{\frac{9}{4}-\frac{ M_\mathcal{S}^2}{H^2}}},
\end{equation} and  $z$ is the spacetime interval between the points $x$ and $y$.

We can calculate this same correlation function in the stochastic theory (\ref{equ:stochastic}) using the techniques laid out in ref.~\cite{Markkanen:2019kpv}.
Assuming that the potential in eq.~(\ref{equ:stochastic}) is the constraint effective potential,
\begin{equation}
    \mathcal{V}(\phi)=\mathcal{V}_\mathcal{C}(\phi)=\frac{1}{2}M_\mathcal{C}^2\phi^2+\frac{1}{4}\lambda_\mathcal{C}\phi^4+\mathcal{O}(\phi^6),
\end{equation}
we find the two-point correlator
\begin{align}
\label{equ:stochcorr}
     \langle \phi(x)\phi(y)\rangle_\text{stoch}
     &=\frac{3H^4}{8\pi^2 M_\mathcal{C}^2}
     \left(1-\frac{9}{8\pi^2}\frac{\lambda_\mathcal{C} H^4}{M_\mathcal{C}^4}\right)
\left(\frac{z}{2}\right)^{-\frac{M_\mathcal{C}^2}{3H^2}
\left(1+\frac{9\lambda_\mathcal{C}H^4}{8\pi^2M_\mathcal{C}^4}
\right)
}+\mathcal{O}(\lambda_\mathcal{C}^2),
\nonumber\\
&=
\frac{3H^4}{8\pi^2M_\mathcal{S}^2}\left(\frac{z}{2}\right)^{-\frac{M_\mathcal{S}^2}{3H^2}}+\mathcal{O}(\lambda^2),
\end{align}
where we used the relations (\ref{equ:CvsS}).
Comparing this to eq.~(\ref{equ:qftcorr}), we see that identifying the potential in the stochastic theory (\ref{equ:stochastic}) with the constraint effective potential correctly reproduces the quantum field theory correlation function at one loop order for light fields.

This is a curious result because strictly speaking, as discussed in section~\ref{Sec:EffPots_cEP}, the constraint effective potential is defined on the four-sphere with Euclidean signature, whereas the stochastic theory involves real time evolution and is therefore only meaningful in the Lorentzian de Sitter spacetime. And even though they both are formulated in terms of an averaged field, in the case of the constraint effective potential, the averaging is done over the four-dimensional sphere, while for the stochastic theory it is over a three-dimensional time slice. There has been some earlier discussion in the literature of a possible link between them~\cite{Rajaraman:2010xd,Beneke:2012kn}, and the explicit agreement we have found for three distinct observables---the one-point probability distribution, the vacuum decay rate, and the long-distance limit of the two-point correlator---provides further evidence for it.

\section{Conclusions\label{Sec:Conclusions}}

In this paper, we have calculated the standard and constraint effective potentials of a real scalar field in de Sitter spacetime explicitly at one-loop order. For light fields, the two potentials differ significantly: the standard effective potential is dominated by infrared contributions, while the constraint effective potential is not. This means that the constraint effective potential can be computed reliably in perturbation theory, unlike the standard one.

This does not make the constraint effective potential more correct or more fundamental. They describe the system from different viewpoints, and the appropriate choice depends on the calculation one wants to perform. Even so, it can be useful to reorganise a calculation so that it uses the constraint effective potential, since it remains perturbative.

As a concrete example, we have presented evidence supporting the earlier conjecture~\cite{Camargo-Molina:2022paw} that the potential entering the Starobinsky–Yokoyama stochastic theory~\cite{Starobinsky:1986fx,Starobinsky:1994bd} is the constraint effective potential. At one-loop order, it reproduces the one-point probability distribution of the field, the long-distance limit of the two-point correlation function, and the vacuum lifetime. This is not yet a complete proof. It would be interesting to test whether this remains true at higher loops and away from the light-field limit, which would require going beyond the Starobinsky–Yokoyama theory to a second-order stochastic formulation~\cite{Cable:2020dke,Cable:2022uwd,Cable:2023gdz}.

Our finding that all these three observables are correctly reproduced also serves as a demonstration of the validity of the Starobinsky-Yokoyama theory. This is non-trivial because the observables refer to very different physical phenomena: The one-point probability distribution is about the equilibrium state of the field on the scale of one Hubble length, the two-point correlation function describes field fluctuations on asymptotically long distances, and the vacuum decay rate is a statement about the occurrence of rare non-linear events. It is remarkable that they are all captured accurately by a simple Langevin equation together with the constraint effective potential.

On the other hand, even though the constraint effective potential can be a useful theoretical tool and eq.~(\ref{equ:pphi}) appears to give it a direct physical interpretation in terms of the one-point probability distribution of the averaged field $\bar\phi$, one should treat it with some caution. This is because $\bar\phi$ is defined on the four-sphere obtained by a Wick rotation, and we are not aware of a way of expressing it in terms of variables defined on the physical de Sitter spacetime.

It should be straightforward to extend our calculations to include fermion and gauge fields. This would allow, for instance, the computation of the constraint (or standard) effective potential for the Standard Model Higgs field. 
Because the constraint effective potential does not suffer from the infrared problem, this can be done using perturbation theory. On the other hand, the constraint effective potential is also well suited for a numerical computation using Monte Carlo simulations~\cite{ORaifeartaigh:1986axd}. In de Sitter spacetime, such simulations must be performed on a four-sphere, which is challenging due to the absence of a regular discretisation. Some progress towards this has been made recently~\cite{Brower:2018szu,Brower:2020jqj}.
Either way, the constraint effective potential provides a promising route of connecting quantum field theories of particle physics to cosmological calculations.

\begin{acknowledgments}
\noindent We would like to thank 
Eliel Camargo-Molina, Mariana Carrillo González, Sebastian Céspedes, Diana L\'opez Nacir and Gonzalo Santa Cruz Moreno for useful discussions. LVGC would like to thank the Abdus Salam Centre for Theoretical Physics (formerly the Theory Group) at Imperial College London, where the idea for this work first took shape, for its hospitality and stimulating environment throughout the years of his studies. AR was supported by the STFC Consolidated Grant ST/X000575/1. LVGC was supported by the Royal Society under the grant reference RE22432.
\end{acknowledgments}

\appendix

\section{Integration constants\label{Sec:Appendix_A}}
\noindent In this Appendix, we compute the mass-parameter-independent term in the effective potential. For that purpose, we study the conformally coupled limit of the system, which is given by $M^{2}\rightarrow 2H^{2}$. This condition is equivalent to demanding that $\nu \rightarrow 1/2$ to leading order in four dimensions. As a result, we arrive at
\begin{equation}
    \mathcal{F}_d\!\left(\tfrac{1}{2}\right)
    = \frac{\mu^{4-d}}{2\Omega_{d}}
    \sum_{n=0}^{\infty}
    \frac{(2n+d-1)\,\Gamma(n+d-1)}{\Gamma(n+1)\,\Gamma(d)}
    \ln\!\left[\left(n+\frac{d}{2}\right)\!\left(n+\frac{d-2}{2}\right)\right].
\end{equation}
As in Section~\ref{Sec:sEP}, we analytically continue the spacetime dimension to non-integer values infinitesimally away from $d=4$, obtaining
\begin{equation}
\label{A.2}
    \mathcal{F}\!\left(\tfrac{1}{2}\right)
    = \frac{
        H^{\,4-\varepsilon}\mu^{\varepsilon}
        \Gamma\!\left(\frac{5-\varepsilon}{2}\right)
    }{
        4\pi^{\frac{5-\varepsilon}{2}}
        \Gamma(4-\varepsilon)
    }
    \sum_{n=0}^{\infty}
    \frac{(2n+3-\varepsilon)\,\Gamma(n+3-\varepsilon)}{\Gamma(n+1)}
    \ln\!\left[\left(n+2-\frac{\varepsilon}{2}\right)\!\left(n+1-\frac{\varepsilon}{2}\right)\right].
\end{equation}
Next, we expand the logarithm as a series in $\varepsilon$,
\begin{equation}
    \ln\!\left[\left(n+2-\tfrac{\varepsilon}{2}\right)\!\left(n+1-\tfrac{\varepsilon}{2}\right)\right]
    = \ln\!\left[(n+2)(n+1)\right]
      - \frac{\varepsilon(2n+3)}{2(n+1)(n+2)}
      + \mathcal{O}(\varepsilon^{2}).
\end{equation}
Substituting this expansion into~(\ref{A.2}) gives
\begin{align}
\label{A.4}
    \mathcal{F}\!\left(\tfrac{1}{2}\right)
    = \frac{
        H^{\,4-\varepsilon}\mu^{\varepsilon}
        \Gamma\!\left(\frac{5-\varepsilon}{2}\right)
    }{
        4\pi^{\frac{5-\varepsilon}{2}}
        \Gamma(4-\varepsilon)
    }
    &\left\{
    \sum_{n=2}^{\infty}
    \frac{(2n-1-\varepsilon)\,\Gamma(n+1-\varepsilon)}{\Gamma(n-1)}
    \ln(n)
    \right.\nonumber\\[4pt]
    &\quad+
    \sum_{n=1}^{\infty}
    \frac{(2n+1-\varepsilon)\,\Gamma(n+2-\varepsilon)}{\Gamma(n)}
    \ln(n)
    \nonumber\\[4pt]
    &\quad\left.
    -\,\varepsilon
    \sum_{n=0}^{\infty}
    \frac{(2n+3-\varepsilon)\,\Gamma(n+3-\varepsilon)}{\Gamma(n+1)}
    \frac{2n+3}{2(n+1)(n+2)}
    + \mathcal{O}(\varepsilon^{2})
    \right\},
\end{align}
where logarithmic properties were used to separate the sums and relabel them so that each term appears as a prefactor multiplying $\ln(n)$. This facilitates rewriting the series in closed form.

To evaluate the sums in~(\ref{A.4}), we use the expansion~\cite{Shore:1979as}
\begin{align}
    \frac{\Gamma(n-\varepsilon)}{\Gamma(n)}
    &= n^{-\varepsilon}
    \Bigg[
        1 + \varepsilon\!\left(
            \frac{1}{2n}
            + \frac{1}{12 n^{2}}
            - \frac{1}{120 n^{4}}
            + \mathcal{O}(n^{-5})
        \right)\nonumber\\
    &\qquad
        + \varepsilon^{2}\!\left(
            \frac{1}{2n}
            + \frac{3}{8 n^{2}}
            + \frac{1}{8 n^{3}}
            + \frac{1}{288 n^{4}}
            + \mathcal{O}(n^{-5})
        \right)
        + \mathcal{O}(\varepsilon^{3})
    \Bigg],
\end{align}
together with the identity
\begin{equation}
    -\zeta'(-s+\varepsilon)
    = \sum_{n=1}^{\infty} \frac{\ln(n)}{n^{-s+\varepsilon}},
\end{equation}
and the analytic continuation of $\zeta'(s)$ to a meromorphic function on $\mathbb{C}$. We then obtain the final result
\begin{align}
    \mathcal{F}\!\left(\tfrac{1}{2}\right)
    &= \frac{11 H^{4}}{480\pi^{2}}
    \Bigg[
        \frac{2}{\varepsilon}
        + \ln\!\left(\frac{4\pi\mu^{2}}{H^{2}}\right)
                - \gamma
        \nonumber\\&~~~~~~~~~~~~~~
        + \frac{90}{11}\ln A
        + \frac{15}{11}\ln(2\pi)
        - \frac{60\zeta'(-3)}{11}
        - \frac{120\zeta'(-2)}{11}
        + \frac{1139}{264}
    \Bigg],
\end{align}
where $A$ denotes the Glaisher--Kinkelin constant. Finally, substituting the above results into the definition of the integration constant, we obtain
\begin{align}
\label{equ:F0}
\begin{split}
     \tilde{\mathcal{F}}&= \frac{1101 H^4}{23040 \pi^2} \left[\frac{2}{\varepsilon}
+ \ln \left( \frac{4\pi\mu^2}{H^2} \right)-  \gamma  + \frac{8160}{367}\ln(2\pi)+\frac{3120\zeta(3)}{1101\pi^2}-\frac{1920\zeta'(-3)}{367} +\frac{3191}{1101}\right].
\end{split}
\end{align}
This contribution exhibits a UV divergence, as indicated by the pole in $2/\varepsilon$. Being field independent, it can be absorbed into local gravitational counterterms (cosmological constant and curvature–squared terms) and into the overall normalisation of the path integral, and therefore has no effect on the field–dependent part of the effective potential. In practice, its divergent and scheme–dependent finite parts are removed by the renormalisation of these local operators.

\end{document}